# An effective slip boundary for micro-structured surfaces containing a second immiscible fluid

Yufan Cao[1], Yixin Zhang[1], Pengyu Lv[1] and Huiling Duan[1] †
(Y.Zhang and Y.Cao contribute equally to this work.)
[1]Department of Mechanics and Engineering Science, College of Engineering, Peking University, Beijing 100871, PR China

## Abstract

Effective slip boundary conditions for flows over periodic micro-structured surfaces containing a secondary immiscible fluid are derived. The primary fluid is in the Cassie state, while the geometries of the micro-structures can be arbitrary. We investigate the impact of the second immiscible fluid on external flow, introducing the effect caused by the viscosity difference between two fluids and the inertia effect of the second fluid into classic Navier slip condition. The effective slip length obtained from our theory for flows over rectangular micro-structures is in good agreement with prior analytical findings. We also apply the theory to mushroom-like micro-structures. The derived effective slip velocity also matches well with two-phase numerical simulations. Implementing the slip boundary conditions on micro-structured surfaces produces external flow fields that are aligned well with the simulation results. Employing a multi-scale homogenization method, we dispose of two-phase flows characterized by strong coupling at the fluid-fluid interface. By introducing the framework of lid-driven cavity approximation, our theory finds practical applications across various scenarios.

## 1. Introduction

There is a strong focus on the research of micro-structured surfaces owing to their superhydrophobic properties[1][3]. These surfaces have been shown to significantly reduce drag resistance in both laminar and turbulent flows[4]-[8]. Typically, micro-structured surfaces prevent liquids from entering the cavities between surface protrusions or grooves, resulting in the entrapment of gas in these cavities. The no-slip boundary condition is no longer satisfied since the flows appear to slip over micro-structured surfaces with entrapped gas[9]. The entrapped gas can be regarded as a second immiscible fluid. Hence, it is imperative to investigate slip boundary conditions on micro-structured surfaces containing a second immiscible fluid. To characterize a slip boundary, Navier[10] first introduced a parameter known as the slip length, which has been widely used in the research of micro-structured surfaces. The slip velocity, denoted as $u_s$, on the surfaces is calculated as follows:

$$u_s|_{\partial B} = \lambda \frac{\partial u}{\partial n}|_{\partial B} \tag{1.1}$$

Here, $n$ is the wall normal direction, and the constant $\lambda$ is referred to as the effective slip length. The effective slip length represents a frictional distance beneath the surface at which the shear flow would extrapolate to zero, and a large slip length indicates a significant slip over the surfaces. Numerous studies have been conducted based on the Navier slip model, including theoretical investigations[11]-[18], experimental research[19][21], and numerical simulations[7][22][23]. Nonetheless, determining the effective slip length on micro-structured surfaces remains a substantial challenge.

Philip[11][12] first analyzed the flows over surfaces with mixed no-slip and no-shear conditions. He provided solutions to a class of mixed boundary value problems pertaining to harmonic and biharmonic fields and determined the transverse and longitudinal effective slip lengths over a period of no-shear stripes. In his theoretical analysis, the liquid-gas interfaces are assumed to be no-shear boundaries, and mixed no-slip and no-shear conditions are applied to a flat surface. Lauga and Stone[13] investigated pressure-driven Stokes flows in a circular pipe with periodically distributed no-shear regions and the effective slip length of these flows is evaluated as a function of the relative width of the no-slip and no-shear regions. The viscosity of the gas trapped within micro-structured surfaces is considerably lower when compared to most liquids, rendering the no-shear conditions a reasonable assumption of the liquid-gas interface. The no-shear conditions are widely used in the subsequent theoretical research on effective slip length over micro-structured surfaces containing a subphase gas, as evidenced by the works of Sbragaglia and Prosperetti[14], Davis and Lauga[16], and Crowdy[17]. However, no-shear conditions imply that the gas trapped in cavities can be neglected, resulting in an infinite slip length at the liquid-gas interface. Furthermore, if the second immiscible fluid trapped in these cavities has a high viscosity, such as oil infused in porous surfaces[34], the no-shear conditions become entirely inappropriate. Therefore, it is crucial to consider the effects of the second immiscible fluid when studying flows over micro-structured surfaces containing it.

Some researchers have abandoned the no-shear conditions and analyzed the effects of the second immiscible fluid. Schönecker and Hardt[24][25] introduced a semi-analytical approach in which they considered the liquid-gas interface, assumed to be flat, as a non-zero shear-stress boundary. This assumption resulted in a non-uniform local slip length distribution across the interface. The effective slip length is evaluated as a function of viscosity contrast and the flow geometry. Asmolov and Vinogradova[26][28] et al. proposed a 'gas cushion model', which assumes that dissipation at the liquid-gas interface is dominated by the shearing of a continuous gas layer. The theoretical treatment grounded in a so-called operator method[28], which establishes a mechanism which transplants the flow in the gas subphase to a local slip boundary condition at the liquid–gas interface. Crowdy[18] provided theoretical insights into scenarios where both the effects of a second immiscible fluid and weak meniscus curvature are considered. He employed Green's second identity and a perturbation method for cases with small and large viscosity contrasts, presenting analytical solutions for rectangular grooves. Ji et al. [33] assumes that the flow within cavities is Stokes flow and established basic equations of the interfacial coupled flow. He deduced an integral relation between the slip velocity and velocity gradient, and the relationship is applicable to two-dimensional cases with the surface geometries of the micro-structured surfaces being rectangular. In general, the effects of the second immiscible fluid are studied by either simplifying two-phase flows into single-phase flows through specific assumptions or by assuming that the flow within cavities is Stokes flow (considering only the viscous effect of the second immiscible fluid). Besides, the geometries of micro-structured surfaces in these studies are typically regular.

The purpose of present paper is to study the effects of the second immiscible fluid on external flows over micro-structured surfaces with arbitrary surface geometries. Effective slip boundaries are generated on these surfaces. We employ a method known as the multi-scale homogenization approach, which is an adjoint-based method. From a perspective based on regions rather than specific values of every place in the field, it presents the possibility of transforming the issues introduced by interfaces into an integral problem within a certain region. Jiménez Bolaños & Vernescu[29] have derived the Navier-slip condition for the Stokes flow over a rough surface using the multi-scale homogenization approach, which adds a first-order corrector term to the no-slip condition of a smooth surface. Zampogna, Magnaudet, and Bottaro[30] extended the classical Navier-slip condition for rough surfaces by introducing a third-order Navier-slip tensor. A further improvement was proposed by Bottaro & Naqvi[31], who aimed for a solution with third-order accuracy. Bottaro[32] introduced adjoint functions to consider nonlinear effects within the microscopic region. In the problem concerned in this paper, the presence of phase interfaces results in stress discontinuities, and the motion of the two-phase fluids is closely coupled with each other through the interface. Besides, the inertia effects that can be introduced by fluids with low viscosity also add nonlinearity to this problem. In this work, we innovatively apply the multiscale homogenization method to multiphase flow problems, transforming the coupled effects caused by interfaces into a more manageable problem of integrating over the single-phase flow field. We also theoretically evaluate the role of inertia effects under different parameter conditions and propose an effective approximation method to obtain the equivalent slip boundary conditions which are practically applicable. Specifically, we introduced adjoint functions for both the liquid in the flow region and the second immiscible fluid in the cavities. We then obtained an integral identity on a chosen surface. Using the homogenization method, we derived effective slip boundaries and investigated the impact of the second immiscible fluid on external flows over micro-structured surfaces, where the surface geometries are arbitrary. In Section 2, we outline the governing equations for the two immiscible fluids and the boundary conditions for the problem. Additionally, we derive the dimensionless parameters and dimensionless equations. In Section 3, we introduce adjoint functions to address the problem and derive an integral identity on a selected surface. Subsequently, we establish two-dimensional and three-dimensional theoretical formulations for the effective slip velocity, considering the effects of the second immiscible fluid. In Section 4, two-phase simulations are conducted to verify the theoretical results. By applying the effective slip boundaries to the surfaces, it yields an external flow field that closely matches the two-phase numerical simulations. In Section 5, we perform a qualitative analysis and a parametric study to identify the critical factors affecting the effects of the second immiscible fluid. In Section 6, practical applications are performed by applying the theoretical results.

## 2. Governing equations and boundary conditions

We consider a flow over a superhydrophobic micro-structured surface with subphase gas as sketched in figure 1. The roughness of the surface is provided by small protrusions distributed on the micro-structured surface. The second immiscible fluid is entrapped in cavities between the protrusions. The representative volume element (RVE) is periodic along wall tangent directions ($\hat{x}_1, \hat{x}_3$) and extends along the wall normal direction $\hat{x}_2$. The fluid and the second fluid in cavities are incompressible with constant density $\rho_1$, $\rho_2$ and viscosity $\mu_1$, $\mu_2$. The velocity and pressure fields are $\hat{u}_i$, $\hat{p}_1$ and $\hat{v}_i$, $\hat{p}_2$. In RVE, two incompressible fluids are governed by Navier-Stokes equations,

$$\rho_1 \frac{\partial \hat{u}_i}{\partial \hat{t}} + \rho_1 \hat{u}_j \frac{\partial \hat{u}_i}{\partial \hat{x}_j} = -\frac{\partial \hat{p}_1}{\partial \hat{x}_i} + \mu_1 \frac{\partial^2 \hat{u}_i}{\partial \hat{x}_j \partial \hat{x}_j}, \qquad (2.1)$$

$$\frac{\partial \hat{u}_i}{\partial \hat{x}_i} = 0, \qquad (2.2)$$

$$\rho_2 \frac{\partial \hat{v}_i}{\partial \hat{t}} + \rho_2 \hat{v}_j \frac{\partial \hat{v}_i}{\partial \hat{v}_j} = -\frac{\partial \hat{p}_2}{\partial \hat{x}_i} + \mu_2 \frac{\partial^2 \hat{v}_i}{\partial \hat{x}_j \partial \hat{x}_j}, \qquad (2.3)$$

$$\frac{\partial \hat{v}_i}{\partial \hat{x}_i} = 0. \qquad (2.4)$$

No-slip boundary conditions are applied to fluid-solid boundaries. The velocity and stress are continuous at the fluid-fluid interface. The curvature of the fluid-fluid interface is assumed to be quite small, so that the surface tension at the interface can be neglected. Therefore, the boundary conditions of governing equations are expressed as

follows:

$$\hat{u}_i = \hat{v}_i, \quad \text{on } \Gamma_d$$

$$-\hat{p}_1 n_i + \mu_1 \left( n_j \frac{\partial \hat{u}_i}{\partial \hat{x}_j} + n_j \frac{\partial \hat{u}_j}{\partial \hat{x}_i} \right) = -\hat{p}_2 n_i + \mu_2 \left( n_j \frac{\partial \hat{v}_i}{\partial \hat{x}_j} + n_j \frac{\partial \hat{v}_j}{\partial \hat{x}_i} \right), \quad \text{on } \Gamma_d$$

$$\hat{u}_i = \hat{v}_i = 0, \quad \text{on } \Gamma_{s1} \text{ and } \Gamma_{s2}$$

$\hat{u}_i, \hat{v}_i, \hat{p}_1, \hat{p}_2$ are periodic along the $\hat{x}_1$ and $\hat{x}_3$ directions. $\quad$ on $\Gamma_2$

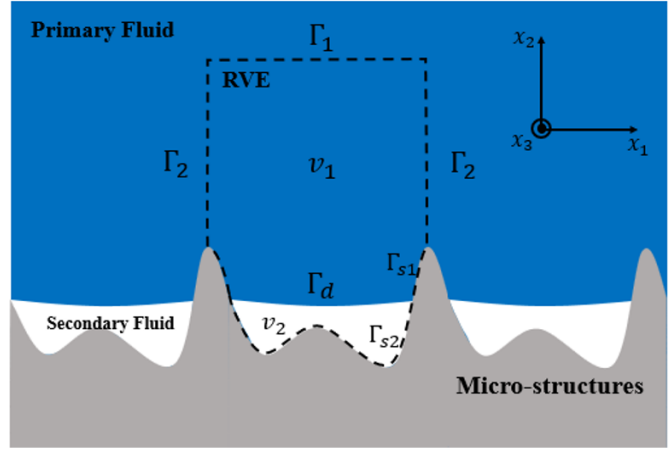

Figure 1. Schematic diagram of a flow over the micro-structured surface with a second immiscible fluid. The black dashed line is the representative volume element (RVE), formed by a primary fluid region $v_1$ and the second fluid region $v_2$. The upper boundary and the periodic boundary of RVE are denoted as $\Gamma_1$ and $\Gamma_2$. The fluid-solid boundaries of the primary fluid and the second fluid are denoted as $\Gamma_{s1}$ and $\Gamma_{s2}$, respectively. The fluid-fluid interface is denoted as $\Gamma_d$. The typical size of the protrusions on the micro-structured surface is l.

To normalize the governing equations in RVE, proper scales should be identified to start the procedure. Since the near-wall flow is driven by outer shear of the flow in the outer region, the microscopic pressure variation within RVE is assumed to be equilibrated by viscous diffusion, so that

$$\Delta P = O\left(\frac{\mu U}{l}\right) \quad (2.5)$$

where $U$ is the characteristic velocity in the surface roughness sublayer (the inner region). We choose $l, U, l/U$ and $\Delta P$ as the length, velocity, time and pressure scales in the near-wall region, so that the dimensionless variables are related to the

dimensional ones through

$$\hat{t} = \frac{l}{U}t, \quad \hat{x}_i = lx_i, \quad \hat{p}_1 = \Delta P p_1, \quad \hat{p}_2 = \Delta P p_2, \quad \hat{u}_i = U u_i, \quad \hat{v}_i = U v_i. \quad (2.6a-f)$$

The dimensionless equations in RVE become

$$Re_1\left(\frac{\partial u_i}{\partial t} + u_j \frac{\partial u_i}{\partial x_j}\right) = -\frac{\partial p_1}{\partial x_i} + \frac{\partial^2 u_i}{\partial x_j \partial x_j}, \qquad (2.7)$$

$$\frac{\partial u_i}{\partial x_i} = 0, \qquad (2.8)$$

$$Re_2\left(\frac{\partial v_i}{\partial t} + v_j \frac{\partial v_i}{\partial v_j}\right) = -\frac{1}{\eta}\frac{\partial p_2}{\partial x_i} + \frac{\partial^2 v_i}{\partial x_j \partial x_j}, \qquad (2.9)$$

$$\frac{\partial v_i}{\partial x_i} = 0, \qquad (2.10)$$

where $Re_1 = \rho_1 U l/\mu_1$, $Re_2 = \rho_2 U l/\mu_2$ and $\eta = \mu_2/\mu_1$. The dimensionless boundary conditions on fluid-fluid interface $\Gamma_d$ become

$$u_i = v_i, \qquad (2.11)$$

$$-p_1 n_i + \left(n_j \frac{\partial u_i}{\partial x_j} + n_j \frac{\partial u_j}{\partial x_i}\right) = -p_2 n_i + \eta \left(n_j \frac{\partial v_i}{\partial x_j} + n_j \frac{\partial v_j}{\partial x_i}\right) \qquad (2.12)$$

where $n_i$ is the normal vector components of the fluid-fluid interface $\Gamma_d$.

In the outer region, the size of large-scale flow structures is denoted as $L$. We consider that the protrusions of the micro-structured surface is sufficiently small, and a parameter $\epsilon$ is introduced as

$$\epsilon = \frac{l}{L} \ll 1. \qquad (2.13)$$

The velocity scale in the outer region is assumed to be $1/\epsilon$ times large than the velocity scale in the RVE. The variables in the outer region are denoted as capital letters, and we have

$$\hat{t} = \frac{L}{U/\epsilon}t = \frac{l}{U}t, \quad \hat{x}_i = LX_i, \quad \hat{p} = P\Delta P, \quad \hat{u}_i = U_i \frac{U}{\epsilon}. \quad (2.14a-d)$$

The chaining of the inner and outer velocity scale is consistent with the fact that a unique time scale $l/U$ appears in the inner and outer region. Besides, the pressure field is continuous across the inner and outer region.

The multi-scale homogenization approach described by Battaro[32] is applied to develop an effective slip boundary over the micro-structured surface. Since the fluid in the inner region evolves two characteristic length scales, we introduce the fast (microscopic) and slow (macroscopic) variables, $x = (x_1, x_2, x_3)$ and $x' = \epsilon(x_1, x_3)$, and the expansions of the fluid variables in the inner region are

$$u_i = u_i^{(0)} + \epsilon u_i^{(1)} + \cdots, \qquad p_1 = p_1^{(0)} + \epsilon p_1^{(1)} + \cdots. \qquad (2.15a-b)$$

where $u_i$ and $p_1$ are functions of $(x_i, x_i', t)$. Furthermore, the operator $\partial/\partial x_i$ is replaced by $\partial/\partial x_i + \epsilon(\partial/\partial x_i')$ based on the chain rule, which means

$$\frac{\partial}{\partial x_i} \rightarrow \frac{\partial}{\partial x_i} + \epsilon \frac{\partial}{\partial x_i'}. \qquad (2.16)$$

By substitute the operators in equations (2.7) and (2.8) and collecting leading order of $\epsilon$, we obtain equations

$$-\frac{\partial p_1^{(0)}}{\partial x_i} + \frac{\partial^2 u_i^{(0)}}{\partial x_j \partial x_j} = 0, \qquad (2.17)$$

$$\frac{\partial u_i^{(0)}}{\partial x_i} = 0 \qquad (2.18)$$

under the assumption that $Re_1$ is at most $O(\epsilon)$. For simplicity, the superscripts of the fluid variables are omitted in the following sections.

## 3. An effective slip boundary for micro-structured surfaces with a second immiscible fluid

Our goal is to develop an effective slip boundary for micro-structured surfaces with a second immiscible fluid. First, a set of adjoint variables $(u_i^+, p_1^+)$ is introduced to fluid region $v_1$. The adjoint variables in fluid region $v_1$ are satisfied with the following equations,

$$\frac{\partial u_i^+}{\partial x_i} = 0, \qquad (3.1)$$

$$-\frac{\partial p_1^+}{\partial x_i} + \frac{\partial^2 u_i^+}{\partial x_j \partial x_j} = 0 \qquad (3.2)$$

with 'no-slip' conditions for $u_i^+$ at the fluid-solid boundary $\Gamma_{s1}$. The adjoint velocity and pressure variables are periodic at the lateral boundaries $\Gamma_2$. We have the freedom to choose boundary conditions at the upper boundary $\Gamma_1$ and the fluid-fluid interface

$\Gamma_d$. Equations (2.17) and (2.18) are multiplied by the adjoint variables $u_i^+$ and $p_1^+$ respectively and integrated in the fluid domain $v_1$, and we have

$$\int_{v_1} \frac{\partial u_i}{\partial x_i} p_1^+ d\Omega = 0, \tag{3.3}$$

$$\int_{v_1} \left( \frac{\partial^2 u_i}{\partial x_j \partial x_j} - \frac{\partial p_1}{\partial x_i} \right) u_i^+ d\Omega = 0. \tag{3.4}$$

From equations (3.1)—(3.4), we could find (see Appendix A)

$$T_1 + \int_{\Gamma_d} n_i p_1^+ u_i d\Sigma - \int_{\Gamma_d} n_i u_i^+ p_1 d\Sigma + \int_{\Gamma_d} n_j u_i^+ \frac{\partial u_i}{\partial x_j} d\Sigma - \int_{\Gamma_d} n_j u_i \frac{\partial u_i^+}{\partial x_j} d\Sigma = 0, \tag{3.5}$$

where $T_1 = \int_{\Gamma_1} n_i p_1^+ u_i d\Sigma - \int_{\Gamma_1} n_i u_i^+ p_1 d\Sigma + \int_{\Gamma_1} n_j u_i^+ \frac{\partial u_i}{\partial x_j} d\Sigma - \int_{\Gamma_1} n_j u_i \frac{\partial u_i^+}{\partial x_j} d\Sigma$ and $n_i$ is the unit normal vector of the fluid domain $v_1$.

Another set of adjoint variables $(v_i^+, p_2^+)$ is introduced to the second fluid domain $v_2$. The governing equations of these adjoint variables are the same as the ones in the fluid domain $v_1$, namely,

$$\frac{\partial v_i^+}{\partial x_i} = 0, \tag{3.6}$$

$$-\frac{\partial p_2^+}{\partial x_i} + \frac{\partial^2 v_i^+}{\partial x_j \partial x_j} = 0 \tag{3.7}$$

with 'no-slip' conditions for $v_i^+$ at the fluid-solid boundary $\Gamma_{s2}$. The boundary conditions at the fluid-fluid interface are unsettled. The equations (2.9) and (2.10) are also multiplied by the adjoint variables $v_i^+$ and $p_2^+$ respectively and integrated in the second fluid domain $v_2$,

$$\int_{v_2} \frac{\partial v_i}{\partial x_i} p_2^+ d\Omega = 0, \tag{3.8}$$

$$Re_2 \int_{v_2} v_i^+ \left( \frac{\partial v_i}{\partial t} + v_j \frac{\partial v_i}{\partial x_j} \right) d\Omega = \int_{v_2} v_i^+ \left( -\frac{1}{\eta} \frac{\partial p_2}{\partial x_i} + \frac{\partial^2 v_i}{\partial x_j \partial x_j} \right) d\Omega. \tag{3.9}$$

From equations (3.6)—(3.9), we could find (see Appendix B)

$$\eta Re_2 S(t) = \int_{\Gamma_d} \eta n_i' p_2^+ v_i d\Sigma - \int_{\Gamma_d} n_i' p_2 v_i^+ d\Sigma + \int_{\Gamma_d} \eta n_j' v_i^+ \frac{\partial v_j}{\partial x_i} d\Sigma - \int_{\Gamma_d} \eta n_j' v_i \frac{\partial v_i^+}{\partial x_j} d\Sigma. \tag{3.10}$$

where $S(t) = \int_{v_2} v_i^+ \left(\frac{\partial v_i}{\partial t} + v_j \frac{\partial v_i}{\partial x_j}\right) d\Omega$ and $n_i'$ is the unit normal vector of the second fluid domain $v_2$.

Since we have the freedom to choose boundary conditions for two sets of adjoint variables at the fluid-fluid interface, the boundary conditions at $\Gamma_d$ are chosen as follows,

$$u_i^+ = v_i^+, \tag{3.11}$$

$$p_1^+ = p_2^+. \tag{3.12}$$

The governing equations of two sets of adjoint variables have the same form, while they are also enforced to be identical at the fluid-fluid interface $\Gamma_d$. Thus, two sets of adjoint variables are identical through the whole flow domain.

Using equations (3.5), (3.7) and (3.10)—(3.12) and boundary conditions of flow variables at the fluid-fluid interface $\Gamma_d$, we could find (see Appendix C)

$$\int_{\Gamma_1} n_i p^+ u_i d\Sigma - \int_{\Gamma_1} n_i u_i^+ p_1 d\Sigma + \int_{\Gamma_1} n_j u_i^+ \frac{\partial u_i}{\partial x_j} d\Sigma - \int_{\Gamma_1} n_j u_i \frac{\partial u_i^+}{\partial x_j} d\Sigma +$$

$$\int_{\Gamma_1} n_j u_i^+ \frac{\partial u_j}{\partial x_i} d\Sigma - \int_{\Gamma_1} n_i u_j \frac{\partial u_i^+}{\partial x_j} d\Sigma + 2(1-\eta) \int_{v_2} \widetilde{S^+} : \widetilde{S}_2 d\Omega - \frac{\rho_2}{\rho_1} Re_1 S(t) = 0. \tag{3.13}$$

where

$$S(t) = \int_{v_2} u_i^+ \left(\frac{\partial v_i}{\partial t} + v_j \frac{\partial v_i}{\partial x_j}\right) d\Omega, \tag{3.14a}$$

$$\widetilde{S^+} = \frac{1}{2}\left(\frac{\partial u_i^+}{\partial x_j} + \frac{\partial u_j^+}{\partial x_i}\right), \tag{3.14b}$$

$$\widetilde{S}_2 = \frac{1}{2}\left(\frac{\partial v_i}{\partial x_j} + \frac{\partial v_j}{\partial x_i}\right). \tag{3.14c}$$

Considering a two-dimensional case, an effective slip boundary is developed over the micro-structured surface and the effect of the second immiscible fluid is analyzed. Applying equation (3.13) to a two-dimensional case, we have

$$\int_{\Gamma_1} (p^+ u_2 - p_1 u_2^+) dl + \int_{\Gamma_1} \left(u_i^+ \frac{\partial u_i}{\partial x_2} - u_i \frac{\partial u_i^+}{\partial x_2}\right) dl + \int_{\Gamma_1} u_i^+ \frac{\partial u_2}{\partial x_i} dl - \int_{\Gamma_1} u_j \frac{\partial u_2^+}{\partial x_j} dl$$

$$+2(1-\eta)\int_{v_2} \widetilde{S^+}:\widetilde{S_2}d\Sigma - \frac{\rho_2}{\rho_1}Re_1 S(t) = 0. \qquad (3.15)$$

The conditions of the adjoint variables at the upper boundary $\Gamma_1$ are chosen as

$$\frac{\partial u_1^+}{\partial x_2} = 1, \quad u_2^+ = 0, \quad at \; x_2 \to y_\infty. \qquad (3.16)$$

The upper boundary conditions are enforced at $y_\infty$ sufficiently far from the surface. The adjoint variables are found to be constant at $y = y_\infty$. The averaged value of variables on upper surface $\Gamma_1$ is defined as

$$\bar{u} = \frac{1}{l_0}\int_{\Gamma_1} u\, dl \qquad (3.17)$$

with $l_0$ the dimensionless length of $\Gamma_1$. Because of the mass conservation and the periodicity of $\Gamma_2$, $\overline{u_2} = 0$ at any $x_2 = y_\infty$.

Provided with all the mentioned constraints, equation (3.15) could be simplified as (see Appendix D)

$$\overline{u_1}|_{y_\infty} = u_1^+ \frac{\overline{\partial u_1}}{\partial x_2}\bigg|_{y_\infty} + \frac{1-\eta}{l_0}\int_{v_2} \frac{\partial u_i^+}{\partial x_j}\left(\frac{\partial v_j}{\partial x_i} + \frac{\partial v_i}{\partial x_j}\right)d\Sigma - \frac{1}{l_0}\frac{\rho_2}{\rho_1} Re_1 S(t). \qquad (3.18)$$

The Taylor expansion

$$\overline{u_1}|_{y_\infty} = \overline{u_1}|_{x_2} + y_\infty \frac{\overline{\partial u_1}}{\partial x_2}\bigg|_{x_2} + \cdots \qquad (3.19)$$

allows us to transfer the effective slip boundary conditions to an arbitrary surface $x = x_2$, which is the effective slip boundary. Thus, the effective slip boundary condition at $x = x_2$ is,

$$\overline{u_1}|_{x_2} = (u_1^+ - y_\infty)\frac{\overline{\partial u_1}}{\partial x_2}\bigg|_{x_2} + \frac{1-\eta}{l_0}\int_{v_2}\frac{\partial u_i^+}{\partial x_j}\left(\frac{\partial v_j}{\partial x_i} + \frac{\partial v_i}{\partial x_j}\right)d\Sigma - \frac{1}{l_0}\frac{\rho_2}{\rho_1} Re_1 S(t). \quad (3.20)$$

The dimensional form of equation (3.20) is

$$\overline{\hat{u}_1}|_{\hat{x}_2} = (u_1^+ - y_\infty)l\frac{\overline{\partial \hat{u}_1}}{\partial \hat{x}_2}\bigg|_{\hat{x}_2} + U\frac{1-\eta}{l_0}\int_{v_2}\frac{\partial u_i^+}{\partial x_j}\left(\frac{\partial v_j}{\partial x_i} + \frac{\partial v_i}{\partial x_j}\right)d\Sigma - \frac{U}{l_0}\frac{\rho_2}{\rho_1} Re_1 S(t). \quad (3.21)$$

The first term in equation (3.20) is the classical Navier-slip condition. The second term represents the viscous effect of the second fluid contributing to the effective slip velocity, which is denoted as $u_{vis} = \frac{1-\eta}{l_0}\int_{v_2}\frac{\partial u_i^+}{\partial x_j}\left(\frac{\partial v_j}{\partial x_i} + \frac{\partial v_i}{\partial x_j}\right)d\Sigma$. The Third term represents the inertia effect of the second fluid, which is denoted as $u_{iner} =$

$-\frac{1}{l_0}\frac{\rho_2}{\rho_1}Re_1 S(t)$. If the density of the second fluid is quite small, the inertia effect could be neglected. In conclusion, the effective slip velocity contains a linear part related to the velocity gradient, and nonlinear parts derived from the viscous and inertia effects of the second fluid.

We can also derive the effective slip boundary conditions over micro-structured surfaces applicable to three-dimensional cases. First, we define two distinct auxiliary problems, whose variables are denoted as $u_{1i}^+, p_1^+$ and $u_{3i}^+, p_3^+$. Two sets of adjoint variables satisfy the same equations (3.1) and (3.2), respectively. The boundary conditions of two sets of adjoint variables at the upper boundary $\Gamma_1$ are as follows,

$$\frac{\partial u_{11}^+}{\partial x_2} = 1, \quad u_{12}^+ = 0, \quad \frac{\partial u_{13}^+}{\partial x_2} = 0 \quad at\ x_2 \to y_\infty \quad (3.22a)$$

$$\frac{\partial u_{31}^+}{\partial x_2} = 0, \quad u_{32}^+ = 0, \quad \frac{\partial u_{33}^+}{\partial x_2} = 1 \quad at\ x_2 \to y_\infty \quad (3.22b)$$

These upper boundary conditions are enforced at $y_\infty$ sufficiently far from the surface, and the adjoint variables are found to be constant here.

Applying equation (3.15) to 3-D cases, we can derive

$$\int_{\Gamma_1}(p_1^+ u_2 - p_1 u_{12}^+)d\Sigma + \int_{\Gamma_1}\left(u_{1i}^+\frac{\partial u_i}{\partial x_2} - u_i\frac{\partial u_{1i}^+}{\partial x_2}\right)d\Sigma + \int_{\Gamma_1}u_{1i}^+\frac{\partial u_2}{\partial x_i}d\Sigma - \int_{\Gamma_1}u_j\frac{\partial u_{12}^+}{\partial x_j}d\Sigma$$
$$+2(1-\eta)\int_{v_2}\widetilde{S_1^+}:\widetilde{S_2}d\Omega - \frac{\rho_2}{\rho_1}Re_1 S_1(t) = 0, \quad (3.23a)$$

$$\int_{\Gamma_1}(p_3^+ u_2 - p_1 u_{32}^+)d\Sigma + \int_{\Gamma_1}\left(u_{3i}^+\frac{\partial u_i}{\partial x_2} - u_i\frac{\partial u_{3i}^+}{\partial x_2}\right)d\Sigma + \int_{\Gamma_1}u_{3i}^+\frac{\partial u_2}{\partial x_i}d\Sigma - \int_{\Gamma_1}u_j\frac{\partial u_{32}^+}{\partial x_j}d\Sigma$$
$$+2(1-\eta)\int_{v_2}\widetilde{S_3^+}:\widetilde{S_2}d\Omega - \frac{\rho_2}{\rho_1}Re_1 S_3(t) = 0, \quad (3.23b)$$

where

$$\widetilde{S_1^+} = \frac{1}{2}\left(\frac{\partial u_{1i}^+}{\partial x_j} + \frac{\partial u_{1j}^+}{\partial x_i}\right), \quad (3.24a)$$

$$S_1(t) = \int_{v_2}u_{1i}^+\left(\frac{\partial v_i}{\partial t} + v_j\frac{\partial v_i}{\partial x_j}\right)d\Omega, \quad (3.24b)$$

$$\widetilde{S_3^+} = \frac{1}{2}\left(\frac{\partial u_{3i}^+}{\partial x_j} + \frac{\partial u_{3j}^+}{\partial x_i}\right), \quad (3.24c)$$

$$S_3(t) = \int_{v_2}u_{3i}^+\left(\frac{\partial v_i}{\partial t} + v_j\frac{\partial v_i}{\partial x_j}\right)d\Omega. \quad (3.24d)$$

In 3-D cases, the averaged value of variables on the upper boundary $\Gamma_1$ is defined as

$$\bar{u} = \frac{1}{\Sigma_0} \int_{\Gamma_1} u d\Sigma, \qquad (3.25)$$

where $\Sigma_0$ is the dimensionless area of $\Gamma_1$. $\overline{u_2}$ equals zero at any $x_2 = y_\infty$ due the mass conservation and periodicity of $\Gamma_2$. Thus, by averaging the variables over the upper boundary of RVE, we can derive (see Appendix D)

$$\overline{u_1}|_{y_\infty} = u_{11}^+ \overline{\frac{\partial u_1}{\partial x_2}}\bigg|_{y_\infty} + u_{13}^+ \overline{\frac{\partial u_3}{\partial x_2}}\bigg|_{y_\infty} + \frac{1-\eta}{\Sigma_0} \int_{v_2} \frac{\partial u_{1i}^+}{\partial x_j}(\frac{\partial v_j}{\partial x_i} + \frac{\partial v_i}{\partial x_j})d\Omega - \frac{1}{\Sigma_0}\frac{\rho_2}{\rho_1} Re_1 S_1(t), (3.26a)$$

$$\overline{u_3}|_{y_\infty} = u_{31}^+ \overline{\frac{\partial u_1}{\partial x_2}}\bigg|_{y_\infty} + u_{33}^+ \overline{\frac{\partial u_3}{\partial x_2}}\bigg|_{y_\infty} + \frac{1-\eta}{\Sigma_0} \int_{v_2} \frac{\partial u_{3i}^+}{\partial x_j}(\frac{\partial v_j}{\partial x_i} + \frac{\partial v_i}{\partial x_j})d\Omega - \frac{1}{\Sigma_0}\frac{\rho_2}{\rho_1} Re_1 S_3(t). (3.26b)$$

Like 2-D cases, the equation (3.26) could be transferred to the effective slip velocity at $\hat{x} = \hat{x}_2$ and expressed in a dimensional form,

$$\overline{\hat{u}_1}|_{\hat{x}_2} = (u_{11}^+ - y_\infty)l\overline{\frac{\partial \hat{u}_1}{\partial \hat{x}_2}}\bigg|_{\hat{x}_2} + u_{13}^+ l\overline{\frac{\partial \hat{u}_3}{\partial \hat{x}_2}}\bigg|_{\hat{x}_2} + U\frac{1-\eta}{\Sigma_0}\int_{v_2}\frac{\partial u_{1i}^+}{\partial x_j}(\frac{\partial v_j}{\partial x_i}+\frac{\partial v_i}{\partial x_j})d\Omega$$

$$- \frac{U}{\Sigma_0}\frac{\rho_2}{\rho_1} Re_1 S_1(t), \qquad (3.27a)$$

$$\overline{\hat{u}_3}|_{\hat{x}_2} = (u_{31}^+ - y_\infty)l\overline{\frac{\partial \hat{u}_1}{\partial \hat{x}_2}}\bigg|_{\hat{x}_2} + u_{33}^+ l\overline{\frac{\partial \hat{u}_3}{\partial \hat{x}_2}}\bigg|_{\hat{x}_2} + U\frac{1-\eta}{\Sigma_0}\int_{v_2}\frac{\partial u_{3i}^+}{\partial x_j}(\frac{\partial v_j}{\partial x_i}+\frac{\partial v_i}{\partial x_j})d\Omega$$

$$- \frac{U}{\Sigma_0}\frac{\rho_2}{\rho_1} Re_1 S_3(t). \qquad (3.27b)$$

## 4. The verification of the effective slip boundary theory

Two-dimensional simulations are conducted to verify the effective slip boundary theory. Two-dimensional flows over micro-structured surfaces with a second immiscible fluid is simulated. As a typical example, water (density $\rho_1 = 10^3 kg \cdot m^{-3}$ and dynamic viscosity $\eta_1 = 10^{-6} m^2 \cdot s^{-1}$) and air (density $\rho_2 = 1.216 kg \cdot m^{-3}$ and dynamic viscosity $\eta_2 = 1.48 \times 10^{-5} m^2 \cdot s^{-1}$) have been adopted in the simulations. The height of the main flow domain L is $500\mu m$, and the width of the RVE $l_0$ is $150\mu m$. The cavity is square with a side length of $b$. The fluid-fluid interface fraction is defined as

$$a = \frac{b}{l_0}. \tag{4.1}$$

The cavity is filled with entrapped gas. Five cases are simulated from a = 0.1 − 0.9 (interval 0.2). It should be noted that the dimensions of cavities have been specified as typical examples. The two-phase simulation where $a = 0.5$ is shown in figure 2. Figure 2(a) is the corresponding RVE where $a = 0.5$. Figure 2(b) shows the phase fraction field and streamlines in a steady state. The lateral boundaries of the flow domain are periodic, and far field conditions are applied at the upper boundary. The fluid-solid boundaries are no-slip. The governing equations are solved using the open-source software OpenFOAM. Selecting the incompressible two-phase solver interFoam, the volume of fluid (VOF) model is adopted to simulate two-phase flows. The driving force of the flows is the pressure difference of the lateral boundaries, specified in fvOptions file of the OpenFOAM case. A mean velocity field $u_1 = 0.05 \, m \cdot s^{-1}$ is added to the primary fluid domain.

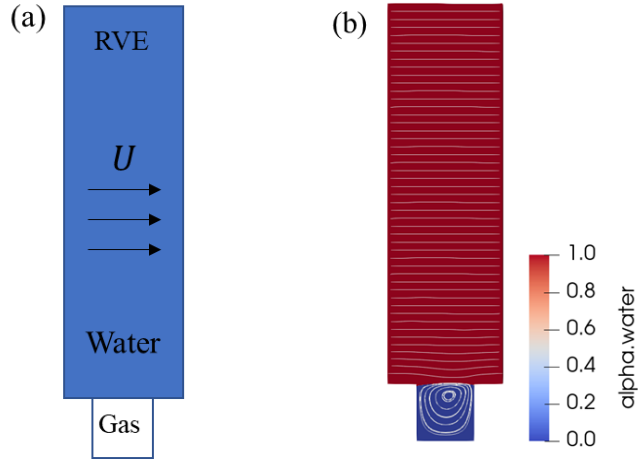

*Figure 2. The simulation of a water flow over a micro-structured surface with entrapped gas where a=0.5. The cavity is filled with entrapped gas. (a)The schematic diagram of RVE, (b) the phase fraction field and streamlines of the two-phase simulation in a steady state.*

To obtain the effective slip velocity, it is necessary to compute the adjoint variables within the RVE by solving their governing equations with appropriate boundary conditions. The governing equations for adjoint variables are

$$\frac{\partial u_i^+}{\partial x_i} = 0, \tag{4.2}$$

$$\frac{\partial^2 u_i^+}{\partial x_j^2} = \frac{\partial p^+}{\partial x_i}. \tag{4.3}$$

The boundary conditions are

$$\frac{\partial u_1^+}{\partial x_2} = 1, \quad u_2^+ = 0 \quad on \ \Gamma_1 \tag{4.4}$$

with 'no-slip' conditions on solid-fluid boundaries and periodic conditions on lateral boundaries. The result of $u_1^+$ where $a = 0.5$ is shown in figure 3(a). Since the governing equations of the adjoint variables are the same as those for incompressible Stokes flow (no convection terms in equations), the 'flow' of adjoint variables resembles Stokes flow. See figure 3(b) for the 'streamlines' in the cavity.

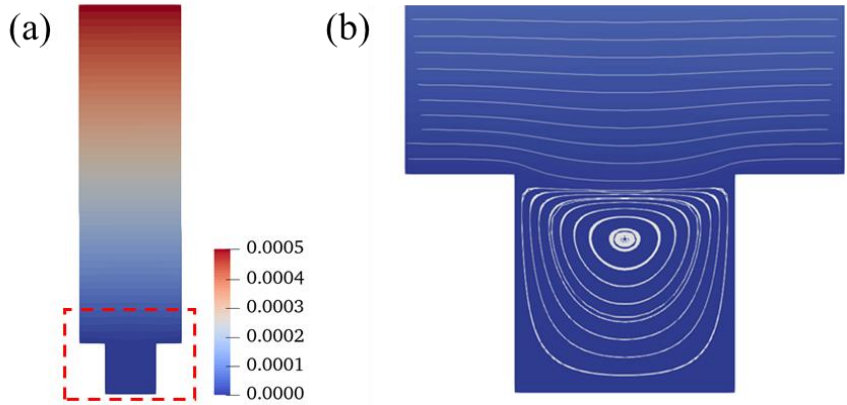

*Figure 3. (a) The result of $u_1^+$ where a=0.5, (b)The 'streamlines' for the adjoint variables in the cavity. The region is corresponding to the red dashed line.*

The effective slip velocities and viscous effects are determined based on the simulation results of adjoint variables and flow variables using Equation (3.20). The effective slip length can be obtained by

$$b_{eff} = \frac{\overline{u_1}}{\overline{\frac{\partial u_1}{\partial x_2}}}. \tag{4.5}$$

To validate our theory, the calculated effective slip lengths with different fluid-fluid interface fraction $a$ and viscosity ratio $\eta$ are compared with the results of Schönecker et al.[24]. Figure 4 shows the behaviors of $b_{eff}$ at changing the fluid-fluid interface fraction $a$. The various dashed lines represent results from Schönecker, corresponding to different viscosity ratios. Among these, $\eta = 0.018$ represents the scenario of water flowing over microstructured surfaces with entrapped gas. The solid line represents

Philip's result, where the fluid-fluid interface is shear-free. The alignment between the effective slip lengths derived from our theory and Schönecker's analytical solutions is quite strong, as evidenced in figure 4.

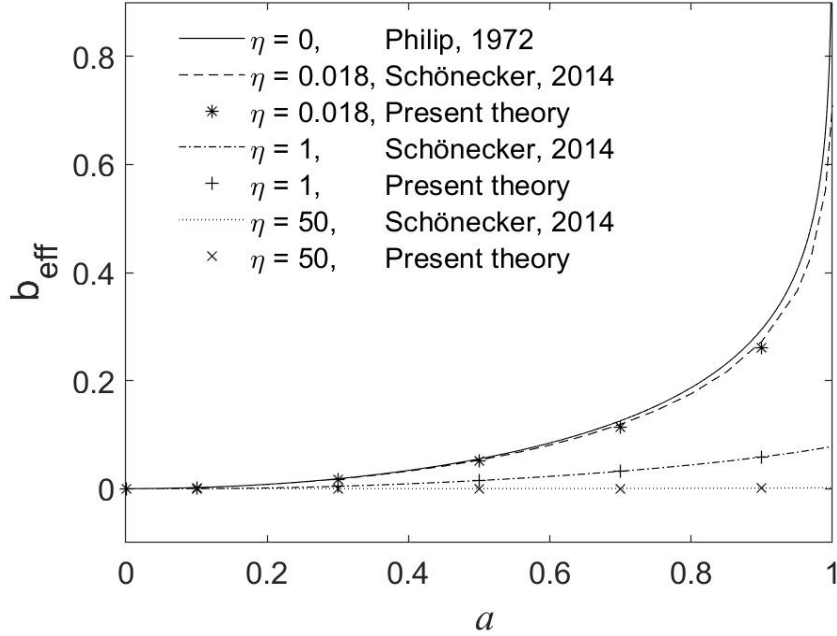

*Figure 4. Effective slip length for various viscosity ratios as a function of the fluid-fluid interface fraction. The dashed lines represent results from Schönecker, while the dots represent the results obtained from our theory.*

The estimated viscous effects and inertia effects are presented in figures 5(a) and 5(b), while the estimated slip velocities are shown in figure 6. The corresponding relative errors are detailed in tables 1 and 2. According to the data magnitude depicted in figures 5(a) and 5(b), the inertia effects can be neglected, which will be discussed further in the following section. As shown in figure 6, the effective slip velocities calculated with our theory based on the simulations match the actual slip velocity veey well. The viscous effects arise from the difference in viscosity between the two fluids. The considerable disparity in viscosity between water and air results in substantial viscous effects, profoundly affecting slip velocities. Additionally, viscous effects and slip velocities can also be estimated through lid-driven cavity flows with constant lid velocities, as described by Schönecker et al.[24]. We conducted simulations of flows within cavities of identical dimensions, driven by a uniform velocity imposed on the lids. Subsequently, we calculated viscous effects and slip velocities using the outcomes of these lid-driven cavity flows. In this work, we regarded this framework of estimation as "lid-driven cavity approximation". The corresponding flow variables are denoted as $U_A, P_A$, and the Equation (3.20) can be written as

$$\overline{u_1}|_{x_2} = (u_1^+ - y_\infty)\frac{\overline{\partial u_1}}{\partial x_2}\bigg|_{x_2} + \frac{1-\eta}{l_0}\int_{v_2}\frac{\partial u_i^+}{\partial x_j}(\frac{\partial U_{A,j}}{\partial x_i} + \frac{\partial U_{A,i}}{\partial x_j})d\Sigma - \frac{1}{l_0}\frac{\rho_2}{\rho_1}Re_1 S_A(t) \quad (4.6)$$

where $S_A(t) = \int_{v_2} u_i^+ \left(\frac{\partial U_{A,i}}{\partial t} + U_{A,j}\frac{\partial U_{A,i}}{\partial x_j}\right)d\Omega$. The results are also illustrated in the figures 5(a) and 6. It is evident that the lid-driven cavity approximation is relatively accurate, rendering it suitable for practical purposes. Further exploration of its practical applications will be covered in Section 6.

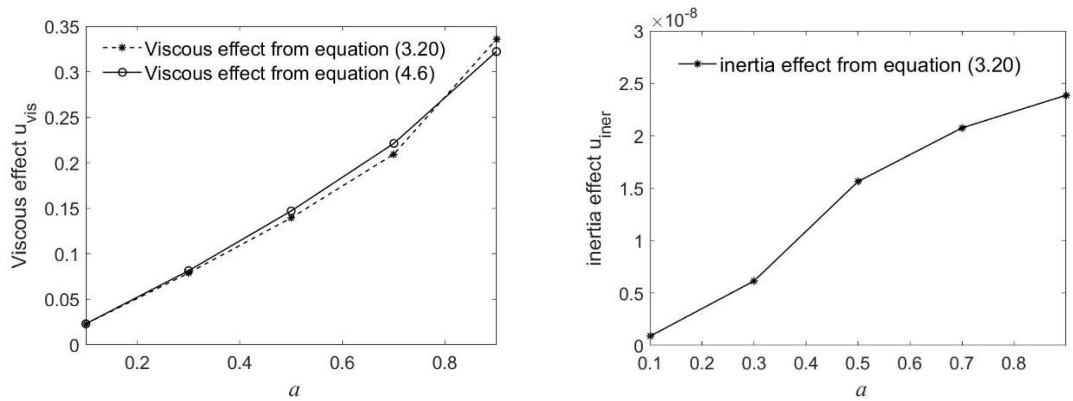

*Figure 5. Viscous effects and inertia effects generated by the theory based on two-phase numerical simulations and the lid-driven cavity approximation. (a)Viscous effects, (b)inertia effects.*

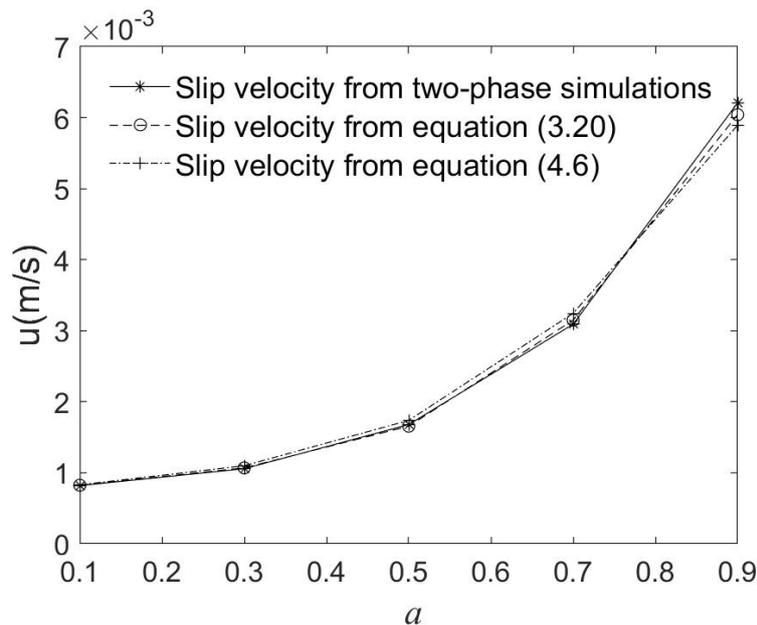

*Figure 6. The real slip velocities obtained from two-phase numerical simulations and slip velocities generated by the theory based on two-phase numerical simulations and the lid-driven cavity approximation.*

*Table 1. Relative Error of Viscous Effects in Lid-driven Cavity Approximation*

| a | 0.1 | 0.3 | 0.5 | 0.7 | 0.9 |
|---|---|---|---|---|---|
| **Relative Error** | 0.0050 | 0.0317 | 0.0545 | 0.0563 | -0.0400 |

**TABLE 1 Relative Error of Viscous Effects in Lid-driven Cavity Approximation**

| a | 0.1 | 0.3 | 0.5 | 0.7 | 0.9 |
|---|---|---|---|---|---|
| **Lid-driven Cavity Approximation** | 0.0136 | 0.0401 | 0.0348 | 0.0495 | -0.0506 |
| **Theory** | 0.0066 | 0.0109 | -0.0137 | 0.0185 | -0.0268 |

**TABLE 2 Relative Error of Slip Velocity from Different Methods**

*Table 2. Relative Error of Slip Velocities from Different Methods*

Specifically, we perform simulations of the external flow utilizing effective slip boundary conditions obtained through our theory when $\eta = 0.018$, $a = 0.5$. These effective slip boundary conditions are generated through the methods discussed above: a two-phase numerical simulation and the lid-driven cavity approximation. We compare the velocity profiles of the external flows with these effective slip boundaries to the average profile obtained from the two-phase flow simulation. The results are presented in Figure 7. The comparison reveals that whether the lid-driven cavity approximation is applied or not, the results based on our theory are aligned well with the outcomes of the direct two-phase flow simulation. This not only validates the accuracy of the theory but also underscores the practicality of the lid-driven cavity approximation.

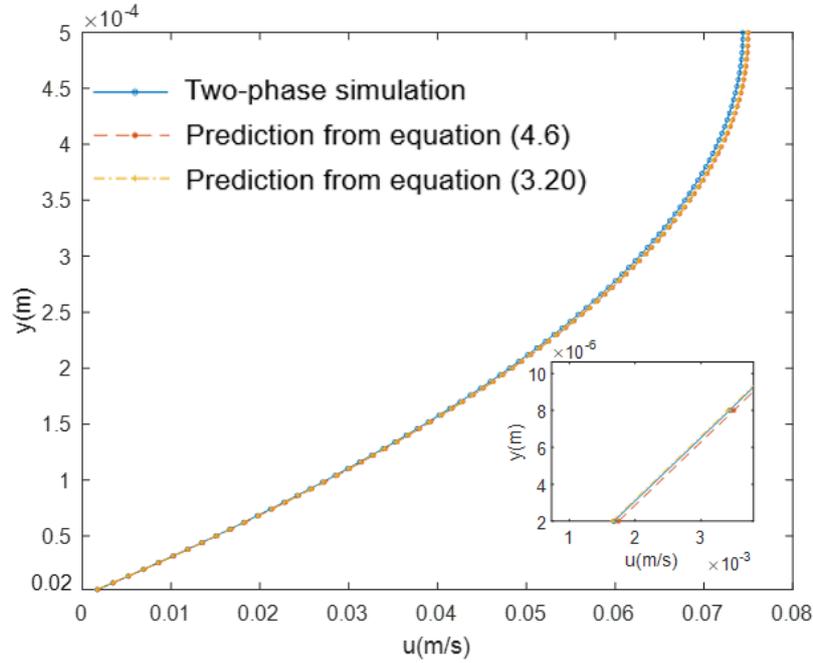

*Figure 7. Velocity Profiles of the external flow. The inset shows the velocity profiles in the near-wall.*

In figure 7, we also present a comparison of differences among the three simulation results specifically in the near-wall regions, as shown in the inset. It becomes clear that the direct simulation results and the predictions based on effective slip boundary conditions computed with the simulation data closely align in the near-wall region, commonly referred to as the linear region in the boundary layer. It is noted that the velocity gradient yielded by all three simulations is identical. Due to the distinctness between the viscous effect evaluated by the lid-driven cavity approximation and the actual value, there exists a disparity between the slip velocity near the wall based on the lid-driven cavity approximation and the direct simulation result. However, this discrepancy remains within an acceptable range.

## 5. Qualitative analysis and parametric study

Our theory can be applied to a wide range of situations. Firstly, we only require a low micro-Reynolds number for the asymptotic expansion of external flows near micro-structured surfaces, making the theory applicable even in high-speed external flow situations where the macro-Reynolds number is relatively high. Secondly, the theory is applicable to situations with different viscosity ratios, such as when the external fluid is a gas phase like air, and the internal fluid is a liquid with higher viscosity, like oil. Thirdly, compared to Bottaro's study[30] of single-phase flows, our results of two-phase flows provide two correction terms, each with a distinct physical significance. In equation (3.20) the first correction term accounts for the viscous effect, while the second correction term represents the inertia effect of the internal fluid. The coefficient of this term is the ratio of the internal fluid density to the external fluid density multiplied by the micro-Reynolds number of the primary flow. Since we have assumed that this micro-Reynolds number is much less than 1 and common gas densities are typically much less than common liquid densities, this term can be neglected when the internal fluid is gas and the external fluid is liquid. This fact also aligns well with the simulation results in Section 4.

Therefore, in the following discussion, we will neglect the second correction term and only take the first correction term into consideration. The first correction term resembles a dissipation rate and is obtained by the dot product of the strain-rate tensor of adjoint functions with the strain-rate tensor of the real flow inside the microstructures. Its coefficient includes $1-\eta$, suggesting that as the dynamic viscosity of the second fluid increases, the reduction of $1-\eta$ tends to decrease the viscous effect. When the ratio of dynamic viscosities between the inner and outer fluids is 1, this term will be 0, indicating that our results revert to Bottaro's results [30] for single-phase flow. Besides, when the dynamic viscosity of the internal fluid tends toward 0, this term will tend toward Philip's theory [11][12].

It is noted that conducting corresponding two-phase flow numerical simulations each time to obtain theoretically accurate effective slip conditions is too costly and not conducive to practical application. Based on the discussion and numerical simulation verification of the "lid-driven cavity approximation" in Section 4, we suggest that in practical scenarios with various microstructure configurations, corresponding computations of lid-driven flows and adjoint functions can be performed instead of direct numerical simulations. This approximation provides relatively accurate results, with significantly lower costs compared to direct two-phase flow simulations, as it just involves conducting lid-driven flow simulations and single-phase Stokes flow simulations for adjoint functions.

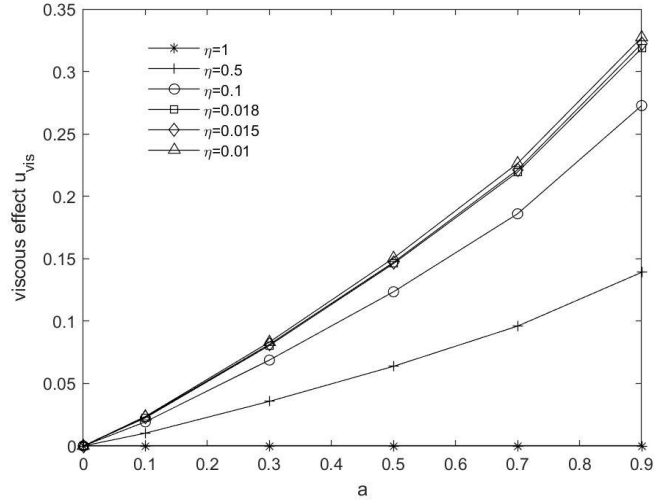

*Figure 8. The viscous effect for various viscosity ratios as function of fluid-fluid interface fraction.*

Since we have verified the feasibility of the lid-driven cavity approximation, we can discuss the relationship between the viscous effect and $\eta$ as well as $a$. Figure 8 shows the viscous effect for various viscosity ratios as function of fluid-fluid interface fraction. For a fixed $\eta$, with the increase of $a$, the viscous effect caused by the second phase fluid gradually enhances, and the growth rate progressively accelerates, which is consistent with the results of Schönecker [25]. For a fixed $a$, with the decrease of $\eta$, the viscous effect caused by the second immiscible fluid also gradually strengthens, but the growth rate gradually slows down, showing a clear trend of convergence, and it can be expected to eventually converge to the situation of Philip's solutions. The curve of $\eta = 0.018$ in this figure corresponds to the situation where the external fluid is water and the fluid inside the microstructures is air. It's evident that employing air as the internal fluid yields a sufficiently strong viscous effect.

## 6. Practical applications

We also apply our theory to a more general situation: water flows over a microstructure with mushroom-like structures and air is trapped in the cavities between the structures. The structure dimensions and flow conditions are depicted in the figure

9(a). A two-phase simulation with the same structure dimensions was conducted, as shown in figure 9(b). By choosing the flow conditions and contact angle, we make the fluid-fluid interface essentially remain flat to satisfy the requirement of our theory. We choose a plane located 0.2 millimeters from the top of the structures as the effective slip boundary. The slip velocity obtained by direct two-phase simulation is 0.009406 m/s, while the effective slip velocities calculated by the simulation results and the lid-driven cavity approximation are 0.00928 m/s and 0.0091 m/s, respectively. The results obtained from our theory closely align with those from the two-phase flow simulation, showing a very good agreement.

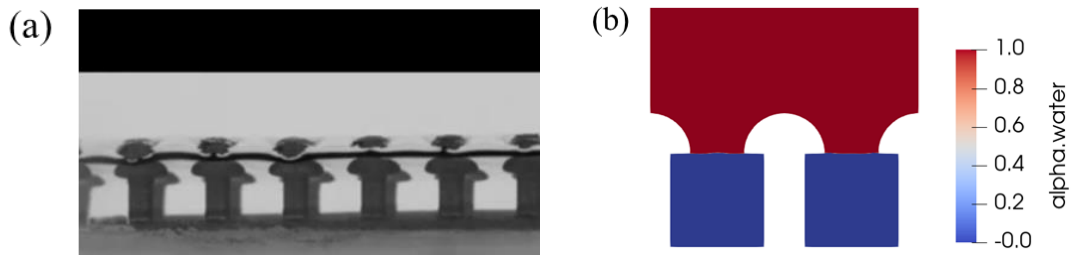

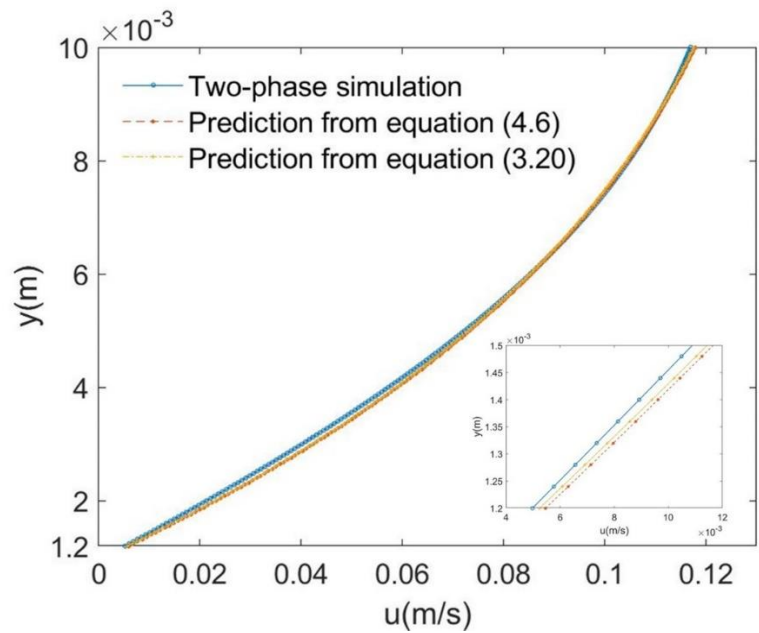

In conclusion, the theoretical results and the lid-driven cavity approximation are validated, and the framework of the approximation can be easily extended to microstructures of different shapes in 2D or 3D situations.

## 7. Conclusion

This paper is concerned with effective slip boundary conditions on micro-structured surfaces containing a second immiscible fluid. Such surfaces widely exist in the natural world and engineering practices, for example, in the simplified modeling of fast-swimming sharkskin or in the simplified modeling of water flow over superhydrophobic surfaces. The study of this flow phenomenon is crucial for the development of drag-reducing biomimetic and metamaterial surfaces, and may also play a significant role in flow control within boundary layers. Research on effective slip boundary conditions on such surfaces can intuitively reveal the effects introduced by the presence of the second fluid. This kind of condition can greatly simplify the numerical simulations of external flow fields in such problems, thus holding significant engineering application value.

Through the classical homogenization method in continuum mechanics and asymptotic expansion, the effective slip boundary conditions for gas-laden surfaces with periodic microstructures, applicable to various geometric configurations, have been derived. The effects of inertia within the microstructures and the differences in viscosity between the inner and outer fluids are accurately modeled and discussed. Rigorous mathematical proofs suggest that, in general application scenarios, the effects of inertia of the second fluid can be neglected. Extensive numerical simulations have been conducted for 2D cases when the shape of grooves is square, and the theoretical results are in excellent agreement with the numerical simulation results. For engineering practicality, we adopted an estimation framework proposed by Schönecker [25] to obtain approximate correction terms using the lid-driven cavity approximation without the need for two-phase flow simulations. The effectiveness of this approximation method has been verified through comparisons with numerical simulations of two-phase flows. Our theory can also be extended to 3D cases, making it applicable to a wider range of scenarios. In comparison to directly assuming the flow of the second fluid as Stokes flow, we have a more rigorous discussion on the inertia effects.

Furthermore, it may be worthwhile to consider the influence of surface tension to some extent. The methods used in this paper can also be extended to two-phase flow problems with temperature fields, and further modeling of inertia effects could potentially yield more accurate results. These areas could be potential directions for future research.

## Appendix A: The derivation from equations (3.1)—(3.4) to Equation (3.5)

Equations (2.17) and (2.18) are multiplied by the adjoint variables $u_i^+$ and $p_1^+$ respectively, and integrated in the fluid domain $v_1$,

$$\int_{v_1} \frac{\partial u_i}{\partial x_i} p_1^+ d\Omega = 0, \tag{A1}$$

$$\int_{v_1} \left( \frac{\partial^2 u_i}{\partial x_j \partial x_j} - \frac{\partial p_1}{\partial x_i} \right) u_i^+ d\Omega = 0. \tag{A2}$$

Gauss Theorem, which is shown below,

$$\int_V a \frac{\partial b}{\partial x_i} dV = \int_{\partial V} abn_i \, d\Sigma - \int_V b \frac{\partial a}{\partial x_i} dV \tag{A3}$$

could be used, and we obtain

$$0 = \int_{v_1} \frac{\partial u_i}{\partial x_i} p_1^+ d\Omega + \int_{v_1} \left( \frac{\partial^2 u_i}{\partial x_j \partial x_j} - \frac{\partial p_1}{\partial x_i} \right) u_i^+ d\Omega$$

$$= \int_{v_1} \frac{\partial (p_1^+ u_i)}{\partial x_i} d\Omega - \int_{v_1} u_i \frac{\partial p_1^+}{\partial x_i} d\Omega + \int_{v_1} u_i^+ \frac{\partial^2 u_i}{\partial x_j \partial x_j} d\Omega - \int_{v_1} u_i^+ \frac{\partial p_1}{\partial x_i} d\Omega$$

$$= \int_{v_1} \frac{\partial (p_1^+ u_i)}{\partial x_i} d\Omega - \int_{v_1} u_i \frac{\partial p_1^+}{\partial x_i} d\Omega + \int_{v_1} \frac{\partial}{\partial x_j} \left( u_i^+ \frac{\partial u_i}{\partial x_j} \right) d\Omega - \int_{v_1} \frac{\partial u_i^+}{\partial x_j} \frac{\partial u_i}{\partial x_j} d\Omega$$

$$- \int_{v_1} \frac{\partial}{\partial x_i} (u_i^+ p_1) d\Omega + \int_{v_1} \frac{\partial u_i^+}{\partial x_i} p_1 d\Omega$$

$$= \int_{\partial v_1} n_i p_1^+ u_i d\Sigma - \int_{\partial v_1} n_i u_i^+ p_1 d\Sigma + \int_{\partial v_1} n_j u_i^+ \frac{\partial u_i}{\partial x_j} d\Sigma - \int_{\partial v_1} n_j u_i \frac{\partial u_i^+}{\partial x_j} d\Sigma$$

$$+ \int_{v_1} u_i \left( \frac{\partial^2 u_i^+}{\partial x_j \partial x_j} - \frac{\partial p_1^+}{\partial x_i} \right) d\Omega + \int_{v_1} \frac{\partial u_i^+}{\partial x_i} p_1 d\Omega. \tag{A4}$$

Noted that the adjoint variables satisfy equations (3.1) and (3.2), substituting them into equation (A4),

$$0 = \int_{\partial v_1} n_i p_1^+ u_i d\Sigma - \int_{\partial v_1} n_i u_i^+ p_1 d\Sigma + \int_{\partial v_1} n_j u_i^+ \frac{\partial u_i}{\partial x_j} d\Sigma - \int_{\partial v_1} n_j u_i \frac{\partial u_i^+}{\partial x_j} d\Sigma. \quad (A5)$$

The boundaries of the domain $v_1$ contain the upper boundary $\Gamma_1$, the periodic boundaries $\Gamma_2$, the fluid-fluid interface $\Gamma_d$ and the fluid-solid boundary $\Gamma_{s1}$. Since 'no-slip' boundary conditions are applied at $\Gamma_{s1}$, and $\Gamma_2$ are periodic boundaries, equation (A5) could be written as,

$$T_1 + \int_{\Gamma_d} n_i p_1^+ u_i d\Sigma - \int_{\Gamma_d} n_i u_i^+ p_1 d\Sigma + \int_{\Gamma_d} n_j u_i^+ \frac{\partial u_i}{\partial x_j} d\Sigma - \int_{\Gamma_d} n_j u_i \frac{\partial u_i^+}{\partial x_j} d\Sigma = 0 \quad (A6)$$

with $T_1 = \int_{\Gamma_1} n_i p_1^+ u_i d\Sigma - \int_{\Gamma_1} n_i u_i^+ p_1 d\Sigma + \int_{\Gamma_1} n_j u_i^+ \frac{\partial u_i}{\partial x_j} d\Sigma - \int_{\Gamma_1} n_j u_i \frac{\partial u_i^+}{\partial x_j} d\Sigma$.

## Appendix B: The derivation from equations (3.6)—(3.9) to equation (3.10)

From equation (3.8)

$$\int_{v_2} \frac{\partial v_i}{\partial x_i} p_2^+ d\Omega = \int_{v_2} \frac{\partial (v_i p_2^+)}{\partial x_i} d\Omega - \int_{v_2} v_i \frac{\partial p_2^+}{\partial x_i} d\Omega = 0,$$

we obtain

$$\int_{\partial v_2} n_i' v_i p_2^+ d\Omega = \int_{v_2} v_i \frac{\partial p_2^+}{\partial x_i} d\Omega. \quad (B1)$$

From equation (3.9),

$$Re_2 \int_{v_2} v_i^+ \left( \frac{\partial v_i}{\partial t} + v_j \frac{\partial v_i}{\partial x_j} \right) d\Omega = Re_2 S(t) = \int_{v_2} v_i^+ \left( -\frac{1}{\eta} \frac{\partial p_2}{\partial x_i} + \frac{\partial^2 v_i}{\partial x_j \partial x_j} \right) d\Omega$$

$$= -\frac{1}{\eta} \int_{v_2} \frac{\partial (p_2 v_i^+)}{\partial x_i} d\Omega + \frac{1}{\eta} \int_{v_2} p_2 \frac{\partial v_i^+}{\partial x_i} d\Omega + \int_{v_2} \frac{\partial}{\partial x_j} \left( v_i^+ \frac{\partial v_i}{\partial x_j} \right) d\Omega - \int_{v_g} \frac{\partial v_i^+}{\partial x_j} \frac{\partial v_i}{\partial x_j} d\Omega$$

$$= -\frac{1}{\eta} \int_{\partial v_2} p_2 v_i^+ n_i' d\Omega + \int_{v_2} \frac{\partial}{\partial x_j} \left( v_i^+ \frac{\partial v_i}{\partial x_j} \right) d\Omega - \int_{v_2} \frac{\partial}{\partial x_j} \left( v_i \frac{\partial v_i^+}{\partial x_j} \right) d\Omega + \int_{v_2} v_i \frac{\partial^2 v_i^+}{\partial x_j \partial x_j} d\Omega$$

$$= -\frac{1}{\eta} \int_{\partial v_2} p_2 v_i^+ n_i' d\Omega + \int_{\partial v_2} n_j' v_i^+ \frac{\partial v_i}{\partial x_j} d\Sigma - \int_{\partial v_2} n_j' v_i \frac{\partial v_i^+}{\partial x_j} d\Sigma + \int_{v_2} v_i \frac{\partial^2 v_i^+}{\partial x_j \partial x_j} d\Omega. \quad (B2)$$

From equations (B1) and (B2), we have

$$Re_2 S(t) = \int_{\partial v_2} n'_i p_2^+ v_i d\Sigma - \frac{1}{\eta} \int_{\partial v_2} n'_i p_2 v_i^+ d\Sigma + \int_{\partial v_2} n'_j v_i^+ \frac{\partial v_j}{\partial x_i} d\Sigma - \int_{\partial v_2} n'_j v_i \frac{\partial v_i^+}{\partial x_j} d\Sigma$$

$$+ \int_{v_2} v_i \left( \frac{\partial^2 v_i^+}{\partial x_j \partial x_j} - \frac{\partial p_2^+}{\partial x_i} \right) d\Omega$$

$$= \int_{\partial v_2} n'_i p_2^+ v_i d\Sigma - \frac{1}{\eta} \int_{\partial v_2} n'_i p_2 v_i^+ d\Sigma + \int_{\partial v_2} n'_j v_i^+ \frac{\partial v_j}{\partial x_i} d\Sigma - \int_{\partial v_2} n'_j v_i \frac{\partial v_i^+}{\partial x_j} d\Sigma. \quad (B3)$$

Thus, from equation (B3) we obtain equation (3.10)

$$\eta Re_2 S(t) = \int_{\Gamma_d} \eta n'_i p_2^+ v_i d\Sigma - \int_{\Gamma_d} n'_i p_2 v_i^+ d\Sigma + \int_{\Gamma_d} \eta n'_j v_i^+ \frac{\partial v_j}{\partial x_i} d\Sigma - \int_{\Gamma_d} \eta n'_j v_i \frac{\partial v_i^+}{\partial x_j} d\Sigma.$$

## Appendix C: The derivation from equations (3.5), (3.7) and (3.10)—(3.12) to equation (3.13)

Summing both sides of equations (3.5) and (3.10),

$$-\eta Re_2 S(t) + T_1 = - \left( \int_{\Gamma_d} \eta n'_i p_2^+ v_i d\Sigma - \int_{\Gamma_d} n'_i p_2 v_i^+ d\Sigma + \int_{\Gamma_d} \eta n'_j v_i^+ \frac{\partial v_j}{\partial x_i} d\Sigma - \int_{\Gamma_d} \eta n'_j v_i \frac{\partial v_i^+}{\partial x_j} d\Sigma \right)$$

$$- \left( \int_{\Gamma_d} n_i p_1^+ u_i d\Sigma - \int_{\Gamma_d} n_i u_i^+ p_1 d\Sigma + \int_{\Gamma_d} n_j u_i^+ \frac{\partial u_i}{\partial x_j} d\Sigma - \int_{\Gamma_d} n_j u_i \frac{\partial u_i^+}{\partial x_j} d\Sigma \right). \quad (C1)$$

The boundary conditions of adjoint variables at fluid-fluid interface $\Gamma_d$ are settled by equations (3.11) and (3.12). In other words, these adjoint variables are identical in the whole domain,

$$u_i^+ = v_i^+, \quad p_1^+ = p_2^+ \quad \text{in } v_1 \cup v_2. \quad (C2)$$

$n'_i$ is the unit normal vector of fluid-fluid interface but contrary to $n_i$. Thus, substituting stress boundary conditions (2.12), we obtain

$$-\eta Re_2 S(t) + T_1 = - \int_{\Gamma_d} \eta n'_i p_1^+ v_i d\Sigma + \int_{\Gamma_d} n'_i p_1^+ u_i d\Sigma + \int_{\Gamma_d} \eta n'_j u_i^+ \frac{\partial v_j}{\partial x_i} d\Sigma$$

$$- \int_{\Gamma_d} n'_j u_i^+ \frac{\partial u_j}{\partial x_i} d\Sigma + \int_{\Gamma_d} \eta n'_j v_i \frac{\partial u_i^+}{\partial x_j} d\Sigma - \int_{\Gamma_d} n'_j u_i \frac{\partial u_i^+}{\partial x_j} d\Sigma. \quad (C3)$$

After simplification of equation (C3) with Gauss Theorem and (3.7), we obtain

$$\eta Re_2 S(t) = T_1 - \int_{\Gamma_1} n_i u_j \frac{\partial u_i^+}{\partial x_j} d\Sigma + \int_{\Gamma_1} n_j u_i^+ \frac{\partial u_j}{\partial x_i} d\Sigma$$
$$+ (1-\eta) \int_{v_2} \frac{\partial u_i^+}{\partial x_j} (\frac{\partial v_j}{\partial x_i} + \frac{\partial v_i}{\partial x_j}) d\Omega. \tag{C4}$$

Noted that from equation (3.5),

$$T_1 = \int_{\Gamma_1} n_i p_1^+ u_i d\Sigma - \int_{\Gamma_1} n_i u_i^+ p_1 d\Sigma + \int_{\Gamma_1} n_j u_i^+ \frac{\partial u_i}{\partial x_j} d\Sigma - \int_{\Gamma_1} n_j u_i \frac{\partial u_i^+}{\partial x_j} d\Sigma. \tag{C5}$$

Substituting equation (C5) into (C4) and considering the periodicity at $\Gamma_2$, we obtain equation (3.13)

$$\int_{\Gamma_1} n_i p^+ u_i d\Sigma - \int_{\Gamma_1} n_i u_i^+ p_1 d\Sigma + \int_{\Gamma_1} n_j u_i^+ \frac{\partial u_i}{\partial x_j} d\Sigma - \int_{\Gamma_1} n_j u_i \frac{\partial u_i^+}{\partial x_j} d\Sigma +$$

$$\int_{\Gamma_1} n_j u_i^+ \frac{\partial u_j}{\partial x_i} d\Sigma - \int_{\Gamma_1} n_i u_j \frac{\partial u_i^+}{\partial x_j} d\Sigma + 2(1-\eta) \int_{v_2} \widetilde{S^+} : \widetilde{S_2} d\Omega - \frac{\rho_1}{\rho_2} Re_1 S(t) = 0$$

with $S(t) = \int_{v_2} u_i^+ \left( \frac{\partial v_i}{\partial t} + v_i \frac{\partial v_i}{\partial x_j} \right) d\Omega$.

## Appendix D: The derivation of equations (3.18) and (3.26)

In 2-D cases, averaging equation (3.15) over upper boundary $\Gamma_1$, we can derive

$$\overline{p^+ u_2} l_0 - \overline{p_1 u_2^+} l_0 + \overline{u_1^+ \frac{\partial u_1}{\partial x_2}} l_0 + \overline{u_2^+ \frac{\partial u_2}{\partial x_2}} l_0 - \overline{u_1 \frac{\partial u_1^+}{\partial x_2}} l_0 - \overline{u_2 \frac{\partial u_2^+}{\partial x_2}} l_0 + \overline{u_1^+ \frac{\partial u_2}{\partial x_1}} l_0$$

$$+ \overline{u_2^+ \frac{\partial u_2}{\partial x_2}} l_0 - \overline{u_1 \frac{\partial u_2^+}{\partial x_1}} l_0 - \overline{u_2 \frac{\partial u_2^+}{\partial x_2}} l_0 + 2(1-\eta) \int_{v_2} \widetilde{S^+} : \widetilde{S_2} d\Sigma - \frac{\rho_2}{\rho_1} Re_1 S(t) = 0. \tag{D1}$$

Considering the conditions of the adjoint variables at the upper boundary $\Gamma_1$, equation (D1) could be simplified as

$$\overline{p^+ u_2} l_0 + \overline{u_1^+ \frac{\partial u_1}{\partial x_2}} l_0 - \overline{u_1} l_0 - \overline{u_2 \frac{\partial u_2^+}{\partial x_2}} l_0 + \overline{u_1^+ \frac{\partial u_2}{\partial x_1}} l_0 - \overline{u_2 \frac{\partial u_2^+}{\partial x_2}} l_0$$

$$+ 2(1-\eta) \int_{v_2} \widetilde{S^+} : \widetilde{S_2} d\Sigma - \frac{\rho_2}{\rho_1} Re_1 S(t) = 0. \tag{D2}$$

Since the adjoint variables are constant at $x_2 = y_\infty$,

$$p^+ \overline{u_2} l_0 + u_1^+ \overline{\frac{\partial u_1}{\partial x_2}} l_0 - \overline{u_1} l_0 - \overline{u_2} \overline{\frac{\partial u_2^+}{\partial x_2}} l_0 + u_1^+ \overline{\frac{\partial u_2}{\partial x_1}} l_0 - \overline{u_2} \overline{\frac{\partial u_2^+}{\partial x_2}} l_0$$

$$+ 2(1-\eta) \int_{v_2} \widetilde{S^+} : \widetilde{S_2} d\Sigma - \frac{\rho_2}{\rho_1} Re_1 S(t) = 0. \tag{D3}$$

Because of the mass conservation and the periodicity of $\Gamma_2$, $\overline{u_2} = 0$. Hence, the equation (D3) is simplified as

$$u_1^+ \overline{\frac{\partial u_1}{\partial x_2}} l_0 - \overline{u_1} l_0 + (1-\eta) \int_{v_2} \overline{\frac{\partial u_i^+}{\partial x_j}\left(\frac{\partial v_j}{\partial x_i} + \frac{\partial v_i}{\partial x_j}\right)} d\Sigma - \frac{\rho_2}{\rho_1} Re_1 S(t) = 0. \quad (D4)$$

Then, we can derive equation (3.18),

$$\overline{u_1}|_{y_\infty} = u_1^+ \overline{\frac{\partial u_1}{\partial x_2}}\bigg|_{y_\infty} + \frac{1-\eta}{l_0} \int_{v_2} \overline{\frac{\partial u_i^+}{\partial x_j}\left(\frac{\partial v_j}{\partial x_i} + \frac{\partial v_i}{\partial x_j}\right)} d\Sigma - \frac{1}{l_0}\frac{\rho_2}{\rho_1} Re_1 S(t).$$

In 3-D cases, we do the same thing to equation (3.23),

$$\overline{p_1^+ u_2}\Sigma_0 - \overline{p_1 u_{12}^+}\Sigma_0 + \overline{u_{11}^+ \frac{\partial u_1}{\partial x_2}}\Sigma_0 + \overline{u_{12}^+ \frac{\partial u_2}{\partial x_2}}\Sigma_0 + \overline{u_{13}^+ \frac{\partial u_3}{\partial x_2}}\Sigma_0 - \overline{u_1 \frac{\partial u_{11}^+}{\partial x_2}}\Sigma_0$$

$$-\overline{u_2 \frac{\partial u_{12}^+}{\partial x_2}}\Sigma_0 - \overline{u_3 \frac{\partial u_{13}^+}{\partial x_2}}\Sigma_0 + \overline{u_{11}^+ \frac{\partial u_2}{\partial x_1}}\Sigma_0 + \overline{u_{12}^+ \frac{\partial u_2}{\partial x_2}}\Sigma_0 + \overline{u_{13}^+ \frac{\partial u_2}{\partial x_3}}\Sigma_0 - \overline{u_1 \frac{\partial u_{12}^+}{\partial x_1}}\Sigma_0$$

$$-\overline{u_2 \frac{\partial u_{12}^+}{\partial x_2}}\Sigma_0 - \overline{u_3 \frac{\partial u_{12}^+}{\partial x_3}}\Sigma_0 + 2(1-\eta)\int_{v_2} \widetilde{S_1^+}:\widetilde{S} d\Omega - \frac{\rho_2}{\rho_1} Re_1 S_1(t) = 0, \quad (D5a)$$

$$\overline{p_3^+ u_2}\Sigma_0 - \overline{p_1 u_{32}^+}\Sigma_0 + \overline{u_{31}^+ \frac{\partial u_1}{\partial x_2}}\Sigma_0 + \overline{u_{32}^+ \frac{\partial u_2}{\partial x_2}}\Sigma_0 + \overline{u_{33}^+ \frac{\partial u_3}{\partial x_2}}\Sigma_0 - \overline{u_1 \frac{\partial u_{31}^+}{\partial x_2}}\Sigma_0$$

$$-\overline{u_2 \frac{\partial u_{32}^+}{\partial x_2}}\Sigma_0 - \overline{u_3 \frac{\partial u_{33}^+}{\partial x_2}}\Sigma_0 + \overline{u_{31}^+ \frac{\partial u_2}{\partial x_1}}\Sigma_0 + \overline{u_{32}^+ \frac{\partial u_2}{\partial x_2}}\Sigma_0 + \overline{u_{33}^+ \frac{\partial u_2}{\partial x_3}}\Sigma_0 - \overline{u_1 \frac{\partial u_{32}^+}{\partial x_1}}\Sigma_0$$

$$-\overline{u_2 \frac{\partial u_{32}^+}{\partial x_2}}\Sigma_0 - \overline{u_3 \frac{\partial u_{32}^+}{\partial x_3}}\Sigma_0 + 2(1-\eta)\int_{v_2} \widetilde{S_3^+}:\widetilde{S_2} d\Omega - \frac{\rho_2}{\rho_1} Re_1 S_3(t) = 0. \quad (D5b)$$

Noted the conditions of the adjoint variables have been enforce at the upper boundary $\Gamma_1$ by equation (3.22), and the adjoint variables are constant here,

$$p_1^+ \overline{u_2}\Sigma_0 + u_{11}^+ \overline{\frac{\partial u_1}{\partial x_2}}\Sigma_0 + u_{13}^+ \overline{\frac{\partial u_3}{\partial x_2}}\Sigma_0 - \overline{u_1}\Sigma_0 - \overline{u_2 \frac{\partial u_{12}^+}{\partial x_2}}\Sigma_0 + u_{11}^+ \overline{\frac{\partial u_2}{\partial x_1}}\Sigma_0$$

$$+u_{13}^+ \overline{\frac{\partial u_2}{\partial x_3}}\Sigma_0 - \overline{u_2 \frac{\partial u_{12}^+}{\partial x_2}}\Sigma_0 + 2(1-\eta)\int_{v_2} \widetilde{S_1^+}:\widetilde{S_2} d\Omega - \frac{\rho_2}{\rho_1} Re_1 S_1(t) = 0, \quad (D6a)$$

$$p_3^+ \overline{u_2}\Sigma_0 + u_{31}^+ \overline{\frac{\partial u_1}{\partial x_2}}\Sigma_0 + u_{33}^+ \overline{\frac{\partial u_3}{\partial x_2}}\Sigma_0 - \overline{u_3}\Sigma_0 - \overline{u_2 \frac{\partial u_{32}^+}{\partial x_2}}\Sigma_0 + u_{31}^+ \overline{\frac{\partial u_2}{\partial x_1}}\Sigma_0$$

$$+u_{33}^+ \overline{\frac{\partial u_2}{\partial x_3}}\Sigma_0 - \overline{u_2 \frac{\partial u_{32}^+}{\partial x_2}}\Sigma_0 + 2(1-\eta)\int_{v_2} \widetilde{S_3^+}:\widetilde{S_2} d\Omega - \frac{\rho_2}{\rho_1} Re_1 S_3(t) = 0. \quad (D6b)$$

Since $\overline{u_2} = 0$ at any $x_2 = y_\infty$,

$$u_{11}^+ \overline{\frac{\partial u_1}{\partial x_2}}\Sigma_0 + u_{13}^+ \overline{\frac{\partial u_3}{\partial x_2}}\Sigma_0 - \overline{u_1}\Sigma_0 + 2(1-\eta)\int_{v_2} \widetilde{S_1^+}:\widetilde{S_2}d\Omega - \frac{\rho_2}{\rho_1}Re_1 S_1(t) = 0, \quad (D7a)$$

$$u_{31}^+ \overline{\frac{\partial u_1}{\partial x_2}}\Sigma_0 + u_{33}^+ \overline{\frac{\partial u_3}{\partial x_2}}\Sigma_0 - \overline{u_3}\Sigma_0 + 2(1-\eta)\int_{v_2} \widetilde{S_3^+}:\widetilde{S_2}d\Omega - \frac{\rho_2}{\rho_1}Re_1 S_3(t) = 0. \quad (D7b)$$

Hence, we derive the equation (3.26)

$$\overline{u_1}|_{y_\infty} = u_{11}^+ \overline{\frac{\partial u_1}{\partial x_2}}\bigg|_{y_\infty} + u_{13}^+ \overline{\frac{\partial u_3}{\partial x_2}}\bigg|_{y_\infty} + \frac{1-\eta}{\Sigma_0}\int_{v_2} \frac{\partial u_{1i}^+}{\partial x_j}(\frac{\partial v_j}{\partial x_i} + \frac{\partial v_i}{\partial x_j})d\Omega - \frac{1}{\Sigma_0}\frac{\rho_2}{\rho_1}Re_1 S_1(t),$$

$$\overline{u_3}|_{y_\infty} = u_{31}^+ \overline{\frac{\partial u_1}{\partial x_2}}\bigg|_{y_\infty} + u_{33}^+ \overline{\frac{\partial u_3}{\partial x_2}}\bigg|_{y_\infty} + \frac{1-\eta}{\Sigma_0}\int_{v_2} \frac{\partial u_{3i}^+}{\partial x_j}(\frac{\partial v_j}{\partial x_i} + \frac{\partial v_i}{\partial x_j})d\Omega - \frac{1}{\Sigma_0}\frac{\rho_2}{\rho_1}Re_1 S_3(t).$$

# References

[1] J. P. Rothstein, Slip on superhydrophobic surfaces, Annu. Rev. Fluid Mech. 42, 89 (2010).

[2] C. Cottin-Bizonne, J. L. Barrat, L. Bocquet, and E. Charlaix, Low-friction flows of liquid at nanopatterned interfaces, Nat. Mater. 2, 237 (2003).

[3] C. H. Choi and C. J. Kim, Large Slip of Aqueous Liquid Flow over a Nanoengineered Superhydrophobic Surface., Phys. Rev. Lett. 96, 066001 (2006).

[4] C. Lee, C. H. Choi, and C. J. Kim, Superhydrophobic drag reduction in laminar flows: a critical review., Exp. Fluids 57, 176 (2016).

[5] J. Ou, B. Perot, and J. P. Rothstein, Laminar drag reduction in microchannels using ultrahydrophobic surfaces., Phys. Fluids 16, 4635 (2004).

[6] H. J. Ling, S. Srinivasan, K. Golovin, G. Mckinley, A. Tuteja, and J. Katz, High-resolution velocity measurement in the inner part of turbulent boundary layers over super-hydrophobic surfaces., J. Fluid Mech. 801, 670 (2016).

[7] T. Min, J. Kim, Effects of hydrophobic surface on skin-friction drag., Physics of Fluids. 16, L55(2004).

[8] E. Aljallis, M. A. Sarshar, R. Datla, et al., Experimental study of skin friction drag reduction on superhydrophobic flat plates in high Reynolds number boundary layer flow., Physics of Fluids. 2013, 025103(2013).

[9] E. Lauga, M. Brenner, H. Stone, Microfluidics: The No-Slip Boundary Condition., Springer Handbook of Experimental Fluid Mechanics, 1219–1240(2007).

[10] C.-L. Navier, M´emoire sur les lois du mouvement des fluides., M´em. Acad. R. Sci. Inst. Fr. 6, 389(1823).

[11] J. R. Philip, Flows satisfying mixed no-slip and shear-free conditions., J. Appl. Math. Phys. 23, 353(1972a).

[12] J. R. Philip, Integral properties of flows satisfying mixed no-slip and shear-free conditions., J. Appl. Math. Phys. 23, 960 (1972b).

[13] E. Lauga and H. A. Stone, Effective slip in pressure-driven Stokes flow., J. Fluid Mech. 489, 55–77 (2003).

[14] M. Sbragaglia and A. Prosperetti, A note on the effective slip properties for microchannel flows with ultrahydrophobic surfaces., Phys. Fluids 19, 043603 (2007).

[15] C. Ybert, C. Barentin, C. Cottin-Bizonne, P. Joseph and L. Bocquet, Achieving large slip with superhydrophobic surfaces: scaling laws for generic geometries. Phys. Fluids 19, 123601(2007).

[16] A. M. J. Davis and E. Lauga, Geometric transition in friction for flow over a bubble mattress., Phys. Fluids 21, 011701 (2009).

[17] D. G. Crowdy, Slip length for longitudinal shear flow over a dilute periodic mattress of protruding bubbles., Phys. Fluids 22, 121703 (2010).

[18] D. G. Crowdy, Perturbation analysis of subphase gas and meniscus curvature effects for longitudinal flows over superhydrophobic surfaces., J. Fluid Mech. 822, 307–326 (2017).

[19] J. Ou, J. P. Rothstein, Direct velocity measurements of the flow past drag-reducing ultrahydrophobic surfaces., Phys. Fluids 17, 103606(2005).

[20] A. Steinberger, C. Cottin-Bizonne, P. Kleimann, and E. Charlaix, High friction on a bubble mattress., Nat. Mater. 6, 665 (2007).

[21] D. Maynes, B. W. Webb and J. Davies, Laminar flow in a microchannel with hydrophobic surface patterned microribs oriented parallel to the flow direction., Phys. Fluids 19, 093603(2007).

[22] C. Cottin-Bizonne, C. Barentin, E. Charlax, L. Bocquet and J. L. Barret, Dynamics of simple liquids at heterogeneous surfaces: molecular-dynamics simulations and hydrodynamic description., Eur. Phys. J. E 15, 427–438(2004).

[23] C.-O. Ng and C. Y. Wang, Stokes shear flow over a grating: Implications for superhydrophobic slip., Phys. Fluids 21, 013602 (2009).

[24] C. Schönecker and S. Hardt, Longitudinal and transverse flow over a cavity containing a second immiscible fluid., J. Fluid Mech. 717, 376–394(2013).

[25] C. Schönecker, T. Baier and S. Hardt, Influence of the enclosed fluid on the flow over a microstructured surface in the Cassie state., J. Fluid Mech. 740, 168–195(2014).

[26] E. S. Asmolov and O. I. Vinogradova, Effective slip boundary conditions for arbitrary one-dimensional surfaces., J. Fluid Mech. 706, 108–117(2012).

[27] T. V. Nizkaya, E. S. Asmolov and O. I. Vinogradova, Flow in channels with superhydrophobic trapezoidal textures., Soft Matt. 9, 11671–11679(2013).

[28] T. V. Nizkaya, E. S. Asmolov and O. I. Vinogradova, Gas cushion model and hydrodynamic boundary conditions for superhydrophobic textures., Phys. Rev. E 90, 043017(2014).

[29] S. J. Bolaños, B. Vernescu, Derivation of the Navier slip and slip length for viscous flows over a rough boundary., Phys. Fluids 29 (5), 057103(2017).

[30] G. A. Zampogna, J. Magnaudet and A. Bottaro, A Generalized slip condition over rough surfaces., J. Fluid Mech. 858, 407–436(2019).

[31] G. A. Zampogna, S.B. Naqvi, J. Magnaudet and A. Bottaro, Compliant riblets: problem formulation and effective macrostructural properties., J. Fluids Struct. 91, 102708(2019).

[32] A. Bottaro, Flow over natural or engineered surfaces: an adjoint homogenization perspective., J. Fluid Mech. 877, P1(2019).

[33] S. Ji, H. Li, Z. Du, P. Lv, H. Duan, Influence of interfacial coupled flow on slip

boundary over a microstructured surface., Phys. Rev. Fluids 8, 054003(2023).

[34] T. S. Wong, S. H. Kang, K. Y. Sindy, E. J. Smythe, B. D. Hatton, A. Grinthal and J. Aizenberg. Bioinspired self-repairing slippery surfaces with pressure-stable omniphobicity., Nature. 477, 443–447 (2011).